\documentclass[12pt,preprint]{aastex}
\usepackage{graphicx}
\usepackage{amsmath,amssymb}

\newcommand{\evm}{{(-)}}
\newcommand{\evz}{{(\circ)}}
\newcommand{\eve}{{(e)}}
\newcommand{\evp}{{(+)}}
\newcommand{\enu}{{(\nu)}}

\newcommand{\gcc}{{\mathrm{g~cm^{-3}}}}

\newcommand{\ddx}[1]{{\frac{{\partial#1}}{\partial x}}}

\usepackage{color}
\begin{document}

\title{On the Piecewise Parabolic Method for Compressible Flow with Stellar Equations of State} 
\author{Michael Zingale and Max P. Katz}
\affil{Dept.\ of Physics and Astronomy, Stony Brook University, Stony Brook, NY 11794-3800}
\email{Michael.Zingale@stonybrook.edu}

\begin{abstract}
The piecewise parabolic method and related schemes are widely used to
model stellar flows.  Several different methods for extending the
validity of these methods to a general equation of state have been
proposed over time, but direct comparisons amongst one-another and
exact solutions with stellar equations of state are not widely
available.  We introduce some simple test problems with exact
solutions run with a popular stellar equation of state and test how
two existing codes with different approaches to incorporating general
gases perform.  The source code for generating the exact solutions is
made available.
\end{abstract}

\section{Introduction}

Many stellar flows are modeled with compressible hydrodynamics
methods, with the piecewise parabolic method
(PPM)~\citep{colellawoodward:1984} being one of the most popular
algorithms.  All extant astrophysical hydrodynamics codes that include
a general equation of state augment the PPM algorithm with additional
information about the energy of the fluid and use Riemann solvers
designed to approximate the wave structure of the solution without
requiring expensive equation of state calls.  However, little
attention in the literature has been given to verifying these
simulation codes directly for the degenerate stellar equation of
state.  While plenty of shock-tube Riemann problems exist for a simple
gamma-law gas (e.g.~\citealt{sod:1978}), these do not test the algorithms
when the gamma of the gas changes dramatically in the problem.  We
introduce some shock tube problems for the general stellar equation of
state and compare the hydrodynamic solution to the exact solution of
the Riemann problem for these tests.  Different codes make different
choices when dealing with a general equation of state, but no
comparisons and little verification has been done on these
implementations.  The basic verification of these methods for stellar
flows is the goal of this paper.

An important diagnostic quantity is the temperature of the gas in the
star.  Degenerate gases are weakly sensitive to temperature---this
creates interesting challenges for hydrodynamics methods.
Traditionally, pressure, not temperature, plays a central role in the
construction of the fluxes through the interfaces in Godunov schemes
for compressible hydrodynamics.  In the piecewise parabolic method
(PPM) implementation \citep{colellawoodward:1984}, the primitive
variables, $\rho, u, p$ are reconstructed as parabolas in each zone
and we then trace under these profiles to find the information that
can reach the interface over the timestep.  Any small errors in the
thermodynamic consistency of the interface state, introduced from the
extrapolation and limiting procedures used in these methods, can lead
to a large error in the temperature calculated from the equation of
state for a degenerate gas.  While not needed in pure hydrodynamic
flows, temperature is an essential quantity for reacting flows and
radiation hydrodynamics.  For this reason, we look at the performance
of the different methods in preventing artifical undershoots and overshoots
in temperature.

In the tests below, we use the publicly-available CASTRO code
\citep{almgren:2010} as our reference and implement the variations as
options in CASTRO.  We also compare to the original
dimensionally-split PPM solver in FLASH~\citep{flash}, that in turn
was based on the PPM implementation in \citet{prometheus}.  While we
note that FLASH has other solvers available, this is the default
solver in FLASH 4.0.  These codes use slightly different
implementations of PPM, based on original method as described by
\citet{colellawoodward:1984} (henceforth CW) and the description in
\citet{millercolella:2002} (henceforth MC).  They also take different
approaches to the extension of the algorithm to a general equation of
state.  Finally, we note a small correction to the scheme used in
FLASH, in Appendix~\ref{app:ge}.

\section{PPM Overview}

The conservative Euler equations appear, in one dimension, as:
\begin{mathletters}
\begin{eqnarray}
\frac{\partial \rho}{\partial t} + \ddx{(\rho u)} &=& 0 \\[0.25em]
\frac{\partial(\rho u)}{\partial t} + \ddx{(\rho u^2)} +\ddx{p} &=& 0 \\[0.25em]
\frac{\partial(\rho E)}{\partial t} + \ddx{(\rho u E + u p)} &=& 0
\end{eqnarray}
\end{mathletters}
where $\rho$ is the density, $u$ is the velocity, $p$ is the
pressure, and $E$ is the total specific energy.  The equations are closed
via an equation of state, 
\begin{equation}
p = p(\rho, e)
\end{equation}
where $e$ is the specific internal energy, $e = E - \frac{1}{2}
u^2$.  For many astrophysical equations of state, temperature ($T$), not 
energy, is an independent variable, and we have
\begin{equation}
p = p(\rho, T), \quad e = e(\rho, T)
\end{equation}
and iterations (usually via the Newton-Raphson method) are done to find the
$T$ that gives the desired $e$, allowing us to then find $p$.

The hydrodynamics equations in terms of the standard set of primitive
variables, $\rho, u, p$, are:
\begin{mathletters}
\begin{eqnarray}
\frac{\partial \rho}{\partial t} + u \ddx{\rho} + \rho\ddx{u} &=& 0 
     \label{eq:rho} \\[0.25em]
\frac{\partial u}{\partial t} + u \ddx{u} + \frac{1}{\rho} \ddx{p} &=& 0 
     \label{eq:u} \\[0.25em]
\frac{\partial p}{\partial t} + u\ddx{p} + \rho c^2 \ddx{u} &=& 0 
     \label{eq:pres}
\end{eqnarray}
\end{mathletters}
where $c$ is the speed of sound defined in terms of the adiabatic
index $\Gamma_1 = \partial (\log p)/\partial(\log \rho) |_s$ as
\begin{equation}
c^2 = {\Gamma_1 p / \rho}
\label{eq:c}
\end{equation}
As most papers deal with a constant-$\gamma$ ideal gas equation of
state, it is important to note that the pressure equation above is
completely general, and can be derived by writing $p = p(\rho, s)$,
and differentiating along streamlines with no entropy sources.  No
assumptions about the constancy of $\gamma$ are needed.

For a general EOS, we need to augment our system with additional
thermodynamic information for the Riemann solver.  CASTRO adds the
evolution of $(\rho e)$ to the system:
\begin{equation}
\frac{\partial (\rho e)}{\partial t} + u\ddx{(\rho e)} + \rho h\ddx{u} = 0 
     \label{eq:e}
\end{equation}
where $h = e + p/\rho$ is the specific enthalpy.  
Alternately, \citet{colellaglaz:1985} (henceforth CG) define $\gamma_e
= p/(\rho e) + 1$ and derive an evolution equation for it.
Differentiating, we find:
\begin{eqnarray}
\frac{D\gamma_e}{Dt} &=& \frac{D}{Dt} \left ( \frac{p}{\rho e} + 1 \right ) 
   = - \frac{p}{(\rho e)^2} \frac{D(\rho e)}{Dt} + \frac{1}{\rho e} \frac{Dp}{Dt} \nonumber \\
   &=& (\gamma_e - 1) (\gamma_e  - \Gamma_1) \frac{\partial u}{\partial x} \label{eq:gammae}
\end{eqnarray}
where we used Eqs.~(\ref{eq:pres}, \ref{eq:c}, \ref{eq:e}) to
simplify.  If we eliminate $\partial u/\partial x$ in favor of $Dp/Dt$
using Eq.~\ref{eq:pres}, then we arrive at the expression from CG.  We
see that for an ideal gas, since $\gamma = \Gamma_1$, we have
$D\gamma/Dt = 0$.

An analysis of this system (see any standard hydrodynamics text,
e.g.\ \citealt{toro:1997}), including one of the energy equations,
shows that there are three characteristic waves that carry jumps in the
variables at the speeds: $\lambda^\evm = u-c$, $\lambda^\evz = u$, and
$\lambda^\evp = u+c$ (with the addition of the $(\rho e)$ or $\gamma_e$
equation, the $u$ eigenvalue becomes degenerate).  We will use these
superscripts, $\evm$, $\evz$, and $\evp$, to distinguish between the
waves in the following discussion.

The PPM algorithm advances the solution through a timestep using
a finite-volume framework.  In each zone, the average value of the
state is  changed by constructing the
fluxes through the interfaces of the zones.  The update appears
as:
\begin{enumerate}
\item Construct limited parabolic profiles of the primitive
  variables, $q_i$, in zone $i$ using the procedure from 
  CW.  This results in a limited parabola, $q(x)$.
 We
 note that alternate limiters exist \citep{colellasekora} and are
 implemented in CASTRO, but we do not consider those here.

\item Integrate under each parabola to find the average state that
  can be carried to each interface by each of the characteristic
  waves. 
  We define an integral $\mathcal{I}_\pm^\enu (q_i)$ which is the
  average of the parabola profile of $q_i$ to the left (`$-$'
  subscript) or right (`$+$' subscript) of the cell for the region
  swept out the wave $\enu$ (see Figure~\ref{fig:trace}).  These
  integrals are:
\begin{mathletters}
\begin{eqnarray}
\mathcal{I}_+^{(\nu)}(q_i) &=& \frac{1}{\sigma^{(\nu)} \Delta x} \int _{x_{i+1/2}
 - \sigma^{(\nu)} \Delta x}^{x_{i+1/2}} q(x) dx \\
\mathcal{I}_-^{(\nu)}(q_i) &=& \frac{1}{\sigma^{(\nu)} \Delta x}
  \int_{x_{i-1/2}}^{x_{i-1/2} + \sigma^{(\nu)} \Delta x} q(x) dx
\end{eqnarray}
\end{mathletters}
with $\sigma^{(\nu)} = |\lambda^{(\nu)}|\Delta t / \Delta x$ (see MC for a discussion
and motivation).

\begin{figure}[t]
\centering
\includegraphics[width=3.75in]{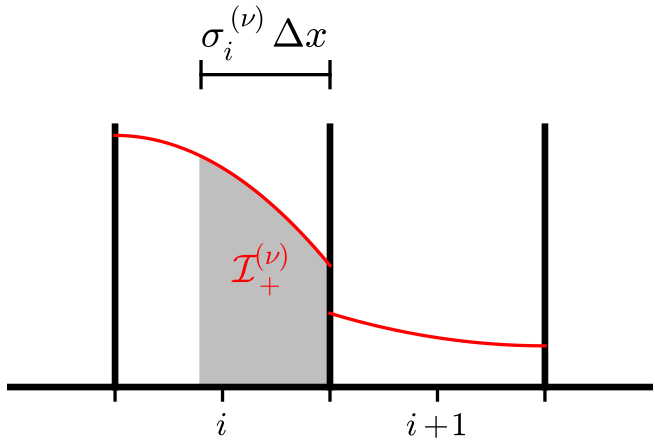}
\caption{\label{fig:trace} Tracing under the parabolic profile to construct
  $\mathcal{I}_+^\enu(q)$ at the $i+1/2$ interface.}
\end{figure}

\item Perform characteristic tracing to build the interface states.
  Following CW and MC, the data in zone $i$ can be used to build the 
  states on the right interface, 
  $q_{i+1/2,L}^{n+1/2}$, and on the left interface,
  $i$, $q_{i-1/2,R}^{n+1/2}$, of zone $i$ (see Figure~\ref{fig:states} for an illustration of the location of each of these states).  These appear as:
  \begin{mathletters}
  \begin{eqnarray}
  q_{i+1/2,L}^{n+1/2} &=& \tilde{q}_+ -
   \sum_{\nu;\lambda^{(\nu)}\ge 0} l_i^{(\nu)} \cdot \left (
        \tilde{q}_+ - \mathcal{I}_+^{(\nu)}(q_i)
       \right ) r_i^{(\nu)} \label{eq:qL} \\
  q_{i-1/2,R}^{n+1/2} &=& \tilde{q}_- - \sum_{\nu; \lambda^\enu \le 0}
   l_i^{(\nu)} \cdot \left ( \tilde{q}_- - \mathcal{I}_-^{(\nu)}(q_i) 
   \right ) r_i^{(\nu)} \label{eq:qR}
  \end{eqnarray}
  \end{mathletters}
  This shows that the same set of eigenvectors and eigenvalues are
  used for these two states.

  \begin{figure}[t]
\centering
\includegraphics[width=3.75in]{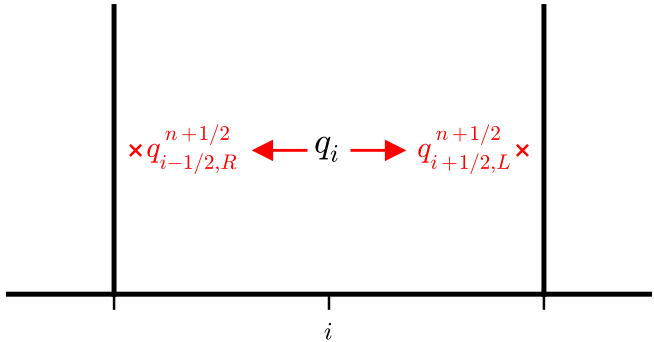}
\caption{\label{fig:states} The two interfaces states constructed from the
  data in zone $i$.}
\end{figure}

  Here, $\tilde{q}_+$ and $\tilde{q}_-$ are the reference states,
  designed to minimize the amount of work done by the characteristic
  tracing (see the discussion in MC).  The key part of this step is
  that the sum (Eqs.~\ref{eq:qL} and \ref{eq:qR}) only includes the
  contributions for waves that are moving toward the interface.  If
  all waves were included, then this would be a no-op (although,
  see the discussion in \citealt{athena} for HLL-type solvers).

  There is a choice in the primitive variable system one can use, and
  therefore, the resulting left and right eigenvectors: $l_i^{(\nu)}$
  and $r_i^{(\nu)}$.  This comes into play with a general gas since
  FLASH and CASTRO use a different ``energy'' variable in the
  eigensystem.  Appendix~\ref{app:eigen} discusses the options and
  gives the form of the eigenvectors, and writes the states
  $q_{i+1/2,\{L,R\}}^{n+1/2}$ for each variables in a notation similar
  to that in CW.

\item Solve the Riemann problem to find the fluxes through each
  interface and convert the unique interface states back to the 
  conserved variables.  There are several different Riemann solvers
  used for non-ideal gases in the literature.  We discuss some of the
  options later.

\item Update the conservative variables using these fluxes.

\end{enumerate}

As a side note, we remark that much of the PPM method (in
one-dimension) is done to third order, but the conversion from
conservative to primitive variables is only second-order accurate, and
this conversion itself can introduce some thermodynamic inconsistency.
\citet{McCorquodale} discuss how to derive a 4th-order accurate
method.  Finally, we note that \citet{colella:1990} and MC described
the extension to multiple dimensions.

\subsection{Variations Among PPM Implementations}

\label{sec:variations}

CASTRO was originally written to closely follow the PPM method as
described in MC, which differs from the original PPM implementation
(CW), and the original dimensionally-split PPM solver implemented in
FLASH in several ways.  Since the original paper~\citep{almgren:2010},
a number of small changes were introduced, which are summarized here.
\begin{itemize}
\item CASTRO originally set the reference state, $\tilde{q}_\pm$, to
  be the cell-center value.  It has been updated to use the prescription
  from CW, which uses the integral under the parabola for the fastest
  wave (if the wave moves toward the interface) or the limit of the
  parabolic interpolant if the wave is not moving toward the
  interface.
  These options are controlled in CASTRO through the {\tt castro.ppm\_reference}
  and {\tt castro.ppm\_reference\_edge\_limit} runtime parameters.

\item In the characteristic projection, CW use $\tau \equiv 1/\rho$
  instead of $\rho$ in the eigensystem (see the form of $\beta^0$ in
  CW Eq.~3.7).  MC use $\rho$.  This is discussed in
  Appendix~\ref{app:CW}, and explored in more detail later.  Note
  however that CW do the parabolic reconstruction of $\rho$, not
  $\tau$.  This choice is controlled in CASTRO by the parameter
  {\tt castro.ppm\_tau\_in\_tracing}.


\item In the characteristic tracing, numerous terms appear of the form
  $\beta \equiv (l \cdot \Delta q)$, where $\Delta q$ is some jump in the
  primitive variable $q$.  CW evaluate these $\beta$'s and any other
  quantities in the interface state construction using the reference
  state (this is equivalent to evaluating the eigenvectors with the
  reference state).  MC
  do not explicitly write out the expansion of the dot product and
  leave things in terms of eigenvectors which are constructed using
  the time-level $n$ data.  We use the CW method here in CASTRO, which
  can be selected through the 
  {\tt castro.ppm\_reference\_eigenvectors}.  

\item All variants described above use flattening to prevent shocks from becoming too
  thin, but there is variation in {\em when} this flattening procedure
  should be applied.  In MC, the edge states $q_l$ and $q_r$ are
  constructed, then the PPM monotonization procedure is applied, and
  finally the flattening is done to each parabola.  In the FLASH split
  PPM solver, the flattening is done on the $q_l$ and $q_r$ {\em
    before} the PPM monotonization.  CW do not seem to indicate which
  of these orderings is preferred.  Both of these do the flattening
  before the parabolas are integrated.  We note that the flattening
  function in CW is written in a different form than in MC, but they
  are analytically equivalent.  CASTRO by default applies the
  flattening after the integrals $\mathcal{I}$ are constructed from
  the parabola, when constructing the final edge states.  These
  differences do not seem to influence the solution much.  The different methods
  can be explored in CASTRO through the parameter {\tt
    castro.ppm\_flatten\_before\_integrals}.  We note that not all
  PPM implementations use flattening---by default ENZO~\citep{enzo}
  has it disabled.  In the tests that follow, we adopt the FLASH
  ordering of flattening and then limiting in CASTRO.
\end{itemize}

Finally, CW use a contact steepening algorithm to keep contact
discontinuities thin.  MC and CASTRO do not implement this, but the
FLASH split PPM solver does and uses it by default.  For the FLASH
runs presented here, we disable contact steepening.  Some authors
(e.g.\ \citealt{athena}) suggest limiting on the characteristic
variables themselves.  This is the default in the latest version of
the FLASH dimensionally-split PPM solver, and we will look at its
influence when we compare with FLASH.

\subsection{Extension to a General EOS}

CW do not address how to extend PPM to a general equation of state.
Most works cite CG as the inspiration for dealing with a general EOS,
in particular, for the Riemann problem, but other variations exist.
CG predict an interface value of $\gamma_e \equiv p/(\rho e) + 1$
using an equation that captures the thermodynamic evolution along
streamlines (similar to Eq.~\ref{eq:gammae}).  We note that CG did not
discuss how to use parabolic reconstruction of the fluid quantities
with a general EOS.  Their $\gamma_e$ does not participate in the
characteristic tracing described above; rather, it is constructed from
the predicted interface value of $p$, using the average $\Gamma_1$ and
$\gamma_e$ on either side of the interface.  Since $\Gamma_1$ does not
explicitly appear in the fluxes, CG argue that taking the cell-average
value for the interface is second-order accurate.

The FLASH dimensionally-split PPM solver reconstructs $\gamma_e$ and
$\Gamma_1$ as parabolas and integrates under the $\lambda^\evz$ wave
to get their interface values (note: this does not appear to be documented
in the FLASH paper).  This differs substantially from CG, and suggests
that they both should obey a hyperbolic PDE.  We show in
Appendix~\ref{app:ge} that it is possible to include $\gamma_e$ in the
characteristic projection and it can actually jump across all three
characteristic waves, not just the $\lambda^\evz$ wave as assumed in
FLASH.  We know of no method to allow for a high-order reconstruction
of $\Gamma_1$ to the interface.  In the construction of the interface
states, FLASH evaluates the Lagrangian sound speed, $C$, using
$\tilde{p}$, $\tilde{\rho}$, but with the cell-centered $\Gamma_1$.
This is correct if we choose not to predict $\Gamma_1$ to the
interfaces using parabolic reconstruction, but potentially
inconsistent with a general EOS if you do.  In CW, $\Gamma_1 =
\gamma_e = \gamma$ was constant, so the issue of what $\Gamma_1$ to
use in $\tilde{C}$ was not discussed.

CASTRO includes an evolution equation for $(\rho e)$, which appears as
an additional hyperbolic PDE in the primitive variable system.  This
is reconstructed and enters into the characteristic tracing in the
same fashion as all the other primitive variables (see
Appendix~\ref{app:MC}).

Solving the Riemann problem exactly for a general equation of state is
expensive (see Appendix~\ref{app}).  Both the FLASH
dimensionally-split PPM solver and CASTRO use an approximate two-shock
Riemann solver.  Here, the left and right waves are assumed to be
shocks and jump conditions are used to link to the state in-between
these waves.  This middle state is traditionally called the
``star''-region.  To describe the thermodynamics of our general gas,
auxiliary information is needed to supplement the primitive variables.
CG use the same evolution equation that predicted the interface value
of $\gamma_e$ to predict the value of $\gamma_e$ in the star-region of
the Riemann problem.  This is an iterative method that converges to
find a value of the nonlinear wavespeeds across the (presumed) shocks
in the Riemann problem.  This is then used to construct the energy in
the star-region.  FLASH uses the parabolically-traced values of
$\gamma_e$ as input to the CG Riemann solver.

CASTRO instead uses the interface values of $(\rho e)$ along with $\rho$, $u$, and $p$
as input to the Riemann problem. 
The default solver in CASTRO follows the procedure in
(\citealt{cgf,bellcolellatrangenstein}) where jump conditions are used to
estimate a value of $(\rho e)$ in the star region.  However the option
exists to use the CG Riemann solver with interface
values of $\gamma_e$ set as:
\begin{equation}
{\gamma_e}_s = \frac{p_s}{(\rho e)_s} + 1
\end{equation}
where $p_s$ and $(\rho e)_s$ are the predicted interface values of $p$
and $(\rho e)$, and $s \in \{l,r\}$ indicates whether we are to the left
or right of the interface.

There are other potential choices for Riemann solvers.  The HLL
solvers~\citep{hll,hlle} use the jump conditions and simple estimates
of the wave speeds to give the fluxes directly, and should work for a
general gas without issue.  The HLLC solver advocated by
Toro~\citep{toro:1997} would need modification for a general gas
because of the construction of the speed of the contact wave.

Each of these methods incorporates information about $e$ in some
fashion.  We note that there is a potential for thermodynamic
inconsistency on the interfaces---the quantities $\rho$, $p$, and
$\gamma_e$ or $(\rho e)$ were brought to the interfaces independent of
one another, subjected to limiting, flattening, and characteristic tracing.  In
fact, they are not independent, and there is an error in how well
these interface states obey the equation of state.

\section{Numerical Tests}

Little, if any, comparison of astrophysical hydrodynamics codes to
exact solutions of the shock tube problem has been done.  Generating
exact solutions for an arbitrary equation of state is
straightforward---these are the exact solutions to the Riemann
problem.  \citet{colellaglaz:1985} outline the procedure to exactly
solve the Riemann problem.  In Appendix~\ref{app} we summarize some of
the implementation details and give the initial conditions for four
problems: a Sod-like problem, a double rarefaction, a strong shock,
and conditions mimicking the edge of an under-resolved star.  The first
three are our stellar EOS analogs to the standard test problems from
\citet{toro:1997}.  The tests presented here are not
exhaustive, but sample some of the flow conditions we might encounter
in large-scale simulations.  We note that $\gamma_e$ is not constant in 
these tests---and in test 4, it varies through its entire valid range.

We use CASTRO as the main code to test our ideas, but we also run the
same tests ``as-is'' with the FLASH dimensionally-split PPM solver
\citep{flash} (version 4.0) just for comparison.  This solver in FLASH
can be thought of as the CW method, performing a reconstruction of
both $\gamma_e$ and $\Gamma_1$.  In the discussion below, we use
``FLASH'' to mean this specific dimensionally-split PPM solver.

CASTRO is primarily a 2- and 3-D code, so we do all the runs in 2-D,
with the transverse direction simply replicating the shock tube
problem.  This means that the transverse flux difference will have no
contribution to the interface states.  For all CASTRO tests, we use
the iterative Riemann solver described in \citet{colellaglaz:1985}
(this is enabled with the CASTRO option {\tt castro.use\_colglaz=1})
unless otherwise specified.

The main driver in FLASH is written with dimensional splitting in
mind, so the PPM solver takes a cycle consisting of two timesteps with
equal $\Delta t$ (swapping the ordering of the directional sweeps in
multiple dimensions) and then re-evaluates $\Delta t$ based on the
flow conditions in the domain.  We run FLASH in 1-D, and to have the
timestepping match CASTRO, we allow FLASH to only take a single
timestep per cycle.  Both codes will modify the last timestep to end
at the time defined by the problem.  This change to the timestepping
is the only change we made to FLASH.

We note that some variation of these results can be expected with
varying CFL number or using an initially small timestep, but we will
not explore these details.  All runs use 128 zones, a CFL number of
$0.8$, and take an initial timestep of $0.1$ of the CFL step.  We also
restrict the maximum increase in $\Delta t$ from one step to the next
to be 10\%.  

Both codes use the same full stellar equation of state described in
\citet{timmesswesty,flash}.  A low temperature floor,
$T_\mathrm{small} = 10^4$~K, is imposed by the equation of state.
It is important to note how this small temperature floor is applied.
During the hydrodynamics solve, we typically enter the EOS with $\rho,
e_\mathrm{want}, X_k$ and ask for $p, T$.  The equation of state most
naturally deals with $\rho, T, X_k$ as inputs, so we must do a
Newton-Raphson iteration to get the energy, $e_\mathrm{want}$ we want.
During the iteration, if $T < T_\mathrm{small}$, then we stop our
iteration and return the thermodynamic state corresponding to the
temperature floor.  The problem is that $e_\mathrm{small} = e(\rho,
T_\mathrm{small}, X_k) \ne e_\mathrm{want}$, so we either break energy
conservation by resetting the hydrodynamic state to $e_\mathrm{small}$
or break thermodynamic consistency by keeping $e_\mathrm{want}$.  For
the simulations presented here, we found that keeping the
thermodynamic consistency is more important to getting a good
temperature field, and this is the procedure used in both codes (this
is the default behavior in FLASH and controlled by the {\tt
  eos\_input\_is\_constant} parameter in the {\tt \&extern} namelist
in CASTRO).





\subsection{Comparison of Riemann Solvers}

Our first comparison is to look at the difference in the solution
between the CG Riemann solver and the solver presented in \citet{cgf}
(henceforth called CGF), which was the default Riemann solver used in
\citet{almgren:2010}.  We focus on tests 1 and 4, since those are the
only tests that show substantial differences.
Figure~\ref{fig:riemann} shows the density, velocity, pressure, and
temperature fields for the two cases together with the exact solution.
Figure~\ref{fig:riemann-error} shows the error against the exact
solution for these two runs.  The temperature field shows the most
differences (amongst the two solvers and against the exact solution).
We see an oscillation in the CGF temperature solution behind the
contact discontinuity and a slightly more pronounced undershoot in the
temperature in the CG solver.  Both are unphysical.  Looking at the
density and pressure errors, we see slightly larger error between the
contact and rarefaction in the CGF case than in the CG case.
Figure~\ref{fig:riemann-error-test4} shows the comparison for test 4.
Here we see differences in the velocity on the right of the domain and
differences in achieving the cool temperature behind the rarefaction.
The CGF gets closer to the temperature minimum.  Going forward, we
use the CG solver exclusively to avoid the oscillations see in test 1,
but these results leave the door open to exploring other Riemann
solvers for general equations of state.

\subsection{Using $\rho$ vs.\ $\tau$ in the Characteristic Tracing}

Here we examine the effect of using $\tau = 1/\rho$ instead of $\rho$
in defining the eigensystem used for the characteristic tracing.  In
essence we are comparing the original CW method to CASTRO's implementation
of the MC method.  Again, only test 1 shows any difference, and again
it is slight.  Figure~\ref{fig:test1-MC-CW} shows the temperature
results for test 1.  The difference here is so slight that it seems
that this variation does not really matter.

\subsection{Comparisons with FLASH}

Finally we summarize by comparing CASTRO's method to FLASH.  In some
sense, this allows us to investigate the differences between using
$\gamma_e$ vs.\ $\rho e$ in the interface states (although, see
Appendix~\ref{app:ge}).  We use both the default options in FLASH
(characteristic limiting) and a run with the characteristic limiting
disabled.  Figures~\ref{fig:summary-test1} through
\ref{fig:summary-test4} show the results.  In test 1, the CASTRO
solution appears to suffer the least amount of temperature undershoot
behind the contact, with the FLASH solution without limiting on the
characteristic variables also looking good, and the default FLASH
having the deepest undershoot.  This suggests that the characteristic
limiting may not always be a good idea, and its use should depend on
what quantities you are looking to optimize.  For tests 2 and 3, there
do not seem to be any major differences---all solvers show the same
features.  For test 4, all of the methods have difficulty resolving
the narrow region between the rarefaction and the contact, and largely
look the same, except for the velocity, where all the methods have
trouble finding the correct velocity at the right edge of the domain.

\section{Discussion}

While the basic PPM algorithm is now 30 years old, there are a large
number of variations in the literature.  We looked at the implementations
in CASTRO and FLASH, and performed verification tests on whether they can accurately model
stellar flows.  The main, simple message of this paper is that it is
possible to do verification of stellar hydrodynamics codes on problems
with a real stellar equation of state, and of course, that one should
be in the habit of doing this.  Part of the inspiration for
this exploration was the common appearance of temperature
``funniness'' such as flooring or oscillations (see, e.g.\ the
comparisons in \citealt{ABRZ:I}) in hydrodynamic flows.  These tests
can be used to help design new methods that optimize for a particular
behavior.

Our tests showed that the different approaches that FLASH and CASTRO
take toward incorporating the real equation of state into the solution
appear reasonable and consistent with one another.  They also showed
that the updates to CASTRO highlighed in \S~\ref{sec:variations}
perform well.  In developing the structure of the eigensystem when
$\gamma_e$ was added, we made a recommendation for the FLASH
dimensionally-split PPM solver to improve the prediction of $\gamma_e$
to the interfaces.  We stress that the tests here are not complete and
much more exploration can be done.

The multidimensional treatment of these methods, and dealing with a
general equation of state, is more complex.  In an unsplit method
(e.g.~\citealt{colella:1990}) the interface states are first predicted
in primitive variable form as if the system were one-dimensional, then
converted back to conservative form where a transverse flux difference
is added.  This makes the interface states ``see'' what is happening
in the transverse directions.  Finally, the states are converted back
to primitive form for the Riemann solve.  This conversion can affect
thermodynamic consistency, especially if it is done without involving
the EOS.  We note that with either $(\rho e)$ or $\gamma_e$ as an
auxiliary variable, a transverse term will need to be incorporated.
This is done in CASTRO currently, and an equation of state call can be
avoided by using the multidimensional pressure equation throughout the
procedure, but a detailed exploration of alternatives may still be
illuminating.

Finally we suggest that future stellar hydrodynamics codes test
against the exact solution to the shock tube problem for the stellar
equation of state.  We make our exact Riemann solver and all of the
CASTRO improvements tested here freely available in the CASTRO code
distribution (the Castro {\tt Sod\_stellar/} problem has all the
necessary inputs).  Similar test problems can be designed for the
nuclear equation of state used in core-collapse supernovae
simulations, although we note that the solution procedure from
\cite{colellaglaz:1985} we follow does not work for a non-convex
equation of state, so modifications will be necessary for a general
nuclear equation of state.


\acknowledgements

This research was supported by NSF award AST-1211563.  We thank Ann
Almgren, John Bell, Alan Calder, Sean Couch, and Dongwook Lee for very
helpful discussions and feedback.

\appendix

\clearpage

\section{\label{app:eigen} Eigenvectors and the Characteristic Projection}

Here we summarize the eigensystems corresponding to different sets of
primitive variables.  We will write the interface state generically
as:
\begin{equation}
\label{eq:app:state}
q_s = \tilde{q}_s - \sum_\nu (l^\enu \cdot \Delta q^\enu_s) r^\enu
\end{equation}
where $s$ refers to the left or right state at an interface and the
sum includes only those waves moving toward that interface.  We
introduce the shorthand for a jump: 
\begin{equation}
\Delta q^\enu_s \equiv \tilde{q}_s - \mathcal{I}_s^\enu(q)
\end{equation}

\subsection{\label{app:MC} Standard primitive variables}

The standard set of primitive variables used in CASTRO for the characteristic
tracing and prediction of the interface states are $q = (\rho, u, p,
\rho e)^\intercal$.  This evolution is governed by Eqs.~\ref{eq:rho}
to \ref{eq:pres} and \ref{eq:e}.  Written in the form
\begin{equation}
q_t + A(q) q_x = 0
\end{equation}
the matrix $A$ takes the form
\begin{equation}
A(q) = \left ( \begin{array}{cccc} u & \rho     & 0      & 0 \\
                                   0 & u        & 1/\rho & 0 \\
                                   0 & \rho c^2 & u      & 0 \\
                                   0 & \rho h   & 0      & u
            \end{array} \right )
\end{equation}
with eigenvalues $\lambda^\evm = u -c$, $\lambda^\evz = u$,
$\lambda^\eve = u$, and $\lambda^\evp = u+c$.  We denote the second
instance of the $u$ eigenvalue with the superscript $e$, and its
associated eigenvector only appears when the primitive variable system
is augmented with $\rho e$.

Expressing the eigenvectors as a matrix, $R = (r^\evm|r^\evz|r^\eve|r^\evp)$ and
$L = (l^\evm|l^\evz|l^\eve|l^\evp)^\intercal$, we
have
\begin{equation}
\renewcommand{\arraystretch}{1.3}
R = \left ( \begin{array}{cccc}
   1      & 1   & 0   & 1   \\
  -c/\rho & 0   & 0   & c/\rho \\
   c^2    & 0   & 0   & c^2 \\
   h      & 0   & 1   & h \end{array} \right )
\qquad
L = \left ( \begin{array}{cccc}
   0 & -{\rho}/{2c} & {1}/{2c^2} & 0 \\
   1 & 0           & -{1}/{c^2} & 0 \\
   0 & 0           & -{h}/{c^2} & 1 \\
   0 & {\rho}/{2c} & {1}/{2c^2} & 0
            \end{array} \right )
\end{equation}
The derivation of these follows from $A r^\enu = \lambda^\enu r^\enu$
and $l^\enu A = \lambda^\enu l^\enu$, with $l^{(i)} \cdot r^{(j)} =
\delta_{ij}$.  Note that since the $u$ eigenvalue is degenerate,
$\Delta q^\evz_s = \Delta q^\eve_s$.


We can now write out the form of the interface states takes by simply
multiplying out the dot products and doing the sum.  We introduce the following
notation (based on CW):
\begin{mathletters}
\begin{eqnarray}
\beta^\evm_s \equiv (l^\evm \cdot \Delta q^\evm_s) &=& 
    \dfrac{\rho}{2c} \left (-\Delta u^\evm_s + \dfrac{\Delta p^\evm_s}{\rho c} \right ) \\
\beta^\evz_s \equiv (l^\evz \cdot \Delta q^\evz_s) &=& 
    \Delta \rho^\evz_s - \frac{\Delta p^\evz_s}{c^2} \\
\beta^\eve_s \equiv (l^\eve \cdot \Delta q^\eve_s) &=& 
    -\frac{h \Delta p^\evz_s}{c^2} + \Delta (\rho e)^\evz_s \\
\beta^\evp_s \equiv (l^\evp \cdot \Delta q^\evp_s) &=& 
    \dfrac{\rho}{2c} \left (\phantom{-}\Delta u^\evp_s + \dfrac{\Delta p^\evp_s}{\rho c} \right ) 
\end{eqnarray}
\end{mathletters}
For these $\beta$'s and those that follow in the other sections, if 
the respective wave is not moving toward the interface, then the
associated $\beta^\enu_s$ is set to 0.  With these definitions, and
Eq.~\ref{eq:app:state}, we have
\begin{mathletters}
\begin{eqnarray}
\rho_s &=& \tilde{\rho} - \left ( \beta^\evm_s + \beta^\evz_s + \beta^\evp_s \right ) \\
u_s    &=& \tilde{u}    - \left ( -\frac{c}{\rho} \beta^\evm_s + \frac{c}{\rho} \beta^\evp_s \right ) \\
p_s    &=& \tilde{p}    - \left ( {c^2} \beta^\evm_s + {c^2} \beta^\evp_s \right ) \\
(\rho e)_s &=& \widetilde{(\rho e)} - \left ( h \beta^\evm_s + \beta^\eve_s + h \beta^\evp_s \right )
\end{eqnarray}
\end{mathletters}

\subsection{\label{app:CW} Specific volume in place of density}

CW consider a system with $\tau \equiv 1/\rho$ replacing $\rho$ in the
primitive variable system and do the characteristic tracing in terms
of this.  We will denote this system as $\mathring{q} = (\tau, u, p,
e)^\intercal$.  Now Eq.~\ref{eq:rho} is modified by substituting
$\rho = \tau^{-1}$ and expanding:
\begin{equation}
\frac{\partial \tau}{\partial t} + 
   u \frac{\partial \tau}{\partial x} -
   {\tau} \frac{\partial u}{\partial x} = 0 
\end{equation}
And our system is
\begin{equation}
\mathring{q}_t + \mathring{A} \mathring{q}_x = 0
\end{equation}
with
\begin{equation}
\mathring{A}(\mathring{q}) = \left ( \begin{array}{cccc}
                                  u  & -\tau    & 0  & 0\\
                                  0  &  u       & \tau & 0 \\
                                  0  & c^2/\tau & u    & 0 \\
                                  0  & p\tau   & 0    & u \end{array} \right )
\end{equation}
This has the same characteristic polynomial, $|\mathring{A} - \lambda
I| = 0$ as $A(q)$, so the eigenvalues are unchanged.  The matrices of eigenvectors
are:
\begin{equation}
\renewcommand{\arraystretch}{1.3}
\mathring{R} = \left ( \begin{array}{cccc}
    1         & 1 & 0 & 1 \\
   c/\tau     & 0 & 0 & -c/\tau \\
  -c^2/\tau^2 & 0 & 0 & -c^2/\tau^2 \\
   -p         & 0 & 1 & -p
  \end{array} \right )
\qquad
\mathring{L} = \left ( \begin{array}{cccc}
    0 & {\tau}/{2c} & -{\tau^2}/{2c^2} & 0 \\
    1 & 0           & {\tau^2}/{c^2}   & 0 \\
    0 & 0           & -{p\tau^2}/{c^2} & 1 \\
    0 & -{\tau}/{2c} & -{\tau^2}/{2c^2} & 0
  \end{array} \right )
\end{equation}


As above, we introduce the notation $\mathring{\beta}^\enu \equiv (\mathring{l}^\enu \cdot \Delta \mathring{q}^\enu )$:
\begin{mathletters}
\begin{eqnarray}
\mathring{\beta}^\evm_s \equiv (\mathring{l}^\evm \cdot \Delta \mathring{q}^\evm_s) &=& 
    \dfrac{1}{2C} \left (\phantom{-}\Delta u^\evm_s - \dfrac{\Delta p^\evm_s}{C} \right ) \\
\mathring{\beta}^\evz_s \equiv (\mathring{l}^\evz \cdot \Delta \mathring{q}^\evz_s) &=& 
    \Delta \tau^\evz_s + \frac{\Delta p^\evz_s}{C^2} \\
\mathring{\beta}^\eve_s \equiv (\mathring{l}^\eve \cdot \Delta \mathring{q}^\eve_s) &=& 
    -\frac{p \Delta p^\evz_s}{C^2} + \Delta e^\evz_s \\
\mathring{\beta}^\evp_s \equiv (\mathring{l}^\evp \cdot \Delta \mathring{q}^\evp_s) &=& 
    \dfrac{1}{2C} \left (-\Delta u^\evp_s - \dfrac{\Delta p^\evp_s}{C} \right ) 
\end{eqnarray}
\end{mathletters}
These $\beta$'s (excluding the `$e$' case) are identical to Eq.~3.7 in CW.
Inserting into Eq.~\ref{eq:app:state} gives:
\begin{mathletters}
\begin{eqnarray}
\tau_s &=& \tilde{\tau}_s - \left ( \mathring{\beta}^\evm_s + \mathring{\beta}^\evz_s + \mathring{\beta}^\evp_s \right ) \label{eq:ts}\\
u_s    &=& \tilde{u}_s    - \left ( C \mathring{\beta}^\evm_s - C \mathring{\beta}^\evp_s \right ) \\
p_s    &=& \tilde{p}_s    - \left ( -{C^2} \mathring{\beta}^\evm_s - {C^2} \mathring{\beta}^\evp_s \right ) \label{eq:ps} \\
e_s &=& \tilde{e}_s - \left ( -p \mathring{\beta}^\evm_s + \mathring{\beta}^\eve_s - p \mathring{\beta}^\evp_s \right )
\end{eqnarray}
\end{mathletters}

If we express $\tilde{\tau} = 1/\tilde{\rho}$ and
$\mathcal{I}_s^\evz(\tau) = 1/\mathcal{I}_s^\evz(\rho)$, then
Eqs.~\ref{eq:ts} to \ref{eq:ps} are identical to CW Eq.~3.6, giving:
\begin{equation}
\rho_s = \left [\frac{1}{\tilde{\rho}_s} - \left ( \mathring{\beta}^\evm_s + \mathring{\beta}^\evz_s + \mathring{\beta}^\evp_s \right ) \right ]^{-1}
\end{equation}
This substitution says that we do the parabolic reconstruction of
$\rho$, but the characteristic projection with $\tau$.  We note that
in evaluating these expressions, CW use $\tilde{C} = \sqrt{\Gamma_1
  \tilde{p} \tilde{\rho}}$ instead of $C$ constructed from the
  time-level n data.

\subsection{\label{app:ge} Using $\gamma_e$}

For a general equation of state, we can combine the ideas of CG and CW
to predict $\gamma_e$ on the interfaces by first reconstructing it as
a parabola.  This is in spirit of what the FLASH dimensionally-split
PPM solver does, but as we'll show here, there is a correction term
that is necessary to account for the jump in $\gamma_e$ across all
waves.

We consider the same system as CW, augmented with the evolution 
equation for $\gamma_e$.  We'll consider the primitive variable
system $\check{q} = (\tau, u, p, \gamma_e )^\intercal$.  Using Eq.~\ref{eq:gammae},
our system can be written as:
\begin{equation}
\check{q}_t + \check{A} \check{q}_x = 0
\end{equation}
with
\begin{equation}
\check{A}(\check{q}) = \left ( \begin{array}{cccc}
                                  u  & -\tau    & 0  & 0\\
                                  0  &  u       & \tau & 0 \\
                                  0  & c^2/\tau & u    & 0 \\
                                  0  & -\alpha  & 0    & u \end{array} \right )
\end{equation}
where we write $\alpha \equiv (\gamma_e - 1)(\gamma_e - \Gamma_1)$ for shorthand.
We notice that the structure and elements of this matrix are identical to that of
$\mathring{A}(\mathring{q})$ with the change $p\tau \rightarrow -\alpha$.  The 
eigenvectors are then easily computable and found as:
\begin{equation}
\renewcommand{\arraystretch}{1.3}
\check{R} = \left ( \begin{array}{cccc}
   1         & 1     & 0     & 1 \\
  c/\tau     & 0     & 0     & -c/\tau \\
 -c^2/\tau^2 & 0     & 0     & -c^2/\tau^2 \\
 \alpha/\tau & 0     & 1     & \alpha/\tau \end{array} \right )
\qquad
\check{L} = \left ( \begin{array}{cccc}
     0 & {\tau}/{2c} & -{\tau^2}/{2c^2} & 0 \\
     1 & 0           &  {\tau^2}/{c^2}   & 0 \\
     0 & 0           & {\alpha\tau}/{c^2} & 1 \\
     0 & -{\tau}/{2c} & -{\tau^2}/{2c^2} & 0 \end{array} \right )
\end{equation}


As above, we introduce the notation $\check{\beta}^\enu \equiv (\check{l}^\enu \cdot \Delta \check{q}^\enu )$.  Because of the similarities to the CW system derived above, we 
see that $\check{\beta}^\evm_s = \mathring{\beta}^\evm_s$, $\check{\beta}^\evz_s = \mathring{\beta}^\evz_s$, and $\check{\beta}^\evp_s = \mathring{\beta}^\evp_s$, and 
\begin{equation}
\check{\beta}^\eve_s \equiv (\check{l}^\eve \cdot \Delta \check{q}^\eve_s) =
    \frac{\alpha \Delta p^\evz_s}{\tau C^2} + \Delta {\gamma_e}^\evz_s 
\end{equation}
Considering only how the $\gamma_e$ state appears in the characteristic tracing, 
we see:
\begin{equation}
{\gamma_e}_s = \tilde{\gamma_e}_s - \check{\beta}^\evm_s \frac{\alpha}{\tau} - \check{\beta}^\eve_s - \check{\beta}^\evp_s \frac{\alpha}{\tau}
\end{equation}
and if we simplify things by taking the reference state, $\tilde{\gamma_e}_s$ to be zero, then we find:
\begin{equation}
{\gamma_e}_s = \mathcal{I}_s^\evz(\gamma_e) + \alpha \left [ 
     -\frac{\check{\beta}^\evm}{\tau} + \frac{\mathcal{I}_s^\evz(p)}{\tau C^2} - \frac{\check{\beta}^\evp}{\tau} \right ]
\end{equation}
We note that with a constant-gamma gas (like an ideal gas), then
$\mathcal{I}_s^\evz(\gamma_e) = \gamma_e$ and $\alpha = 0$, as
expected.  Written as above, we see that the $\gamma_e$ interface
state can be viewed as the average of $\gamma_e$ under the
reconstructed parabola over the region traced by the $\lambda^\evz =
u$ wave plus a correction term that is proportional to $\alpha$.  In
FLASH, to the best of our knowledge, only the first term is present in
the dimensionally-split PPM solver.

\section{\label{app} Exact Riemann Solution with a Degenerate Gas}

To test the variations on the standard PPM method we need a test
problem with an exact solution for the general stellar equation of
state.  CG (section 1) describe how to exactly solve the Riemann
problem for a general equation of state.  As with a gamma-law gas,
across the left and right waves, different functions connect the
left/right state to the star state, and the resulting equation for
$p_\star$ needs to be solved via an iterative procedure.  However,
unlike the gamma-law gas, one cannot write down a closed-form
expression for either the rarefaction or shock cases.  For the
rarefaction, a system of ODEs must be integrated, while for the shock,
a root-finding procedure operates on the Rankine-Hugoniot jump
conditions.

Note that in the following, we assume a constant composition.  Adding
a composition jump at the interface is straightforward, as the
composition will only change across the contact wave (this follows
from the structure of the right eigenvectors).
Below we summarize the solution procedure from CG, filling in 
some implementation details:
\begin{itemize}

\item {\em Initial guess:}  We need an initial guess for $p_\star$.
We use the two-shock approximation from \citet{toro:1997}, Eq. 9.42:
\begin{equation}
  p_\star = [(W_r p_l + W_l p_r) + W_l W_r (u_l - u_r)]/(W_l + W_r)
\end{equation}
with the initial guess for the wave speeds taken to be the Lagrangian
sound speed,  $W_s = \sqrt{{\Gamma_1}_s p_s \rho_s}$

\item {\em $p_\star$ iteration loop:} We use perform the Newton
  iteration described in CG.  For $s = \{l,r\}$, we apply the shock
  jump conditions if $p_\star > p_s$, and use the Riemann invariants
  for a rarefaction otherwise.  In each case, our goal is to find $Z_s
  \equiv |dp_\star/du_{\star,s}|$ and $W_s$.

 \begin{itemize}
   \item {\em Shock solution:} For a shock, we solve the Rankine-Hugoniot
   jump conditions.  This takes the form (see CG Eq.~12):
   \begin{equation}
   f(W_s) = W_s^2 \left \{ e \left (p_\star, 
           \left [\frac{1}{\rho_s} - \frac{p_\star - p_s}{W_s^2} \right ]^{-1}
         \right ) - e_s \right \}  - \frac{1}{2} (p_\star^2 - p_s^2)
   \end{equation}
   with the solution corresponding to $f(W_s) = 0$.  Here, we used the mass
   Rankine-Hugoniot condition to express the density in the EOS function
   in terms of $p_\star$ and $W_s$:
   \begin{equation}
   \rho_\star = \left [\frac{1}{\rho_s} - \frac{p_\star - p_s}{W_s^2} \right ]^{-1}
   \end{equation}
   We solve this via
   Newton iteration, using the derivative:
   \begin{equation}
   f^\prime(W_s) = 2 W_s 
       \left \{ e \left (p_\star, \rho_\star \right ) - e_s \right \} - 
       \frac{2}{W_s} \left . 
          \frac{\partial e}{\partial \rho} \right |_p (p_\star - p_s) \rho_\star^2
   \end{equation}
   We iterate until $|f/f^\prime| < \epsilon W_s$, where $\epsilon$ is
   a small tolerance.  We evaluate that thermodynamic derivative by
   expressing our EOS as $e(\rho,T(\rho,p))$:
   \begin{equation}
   \left . \frac{\partial e}{\partial \rho} \right |_p  =
         \left . \frac{\partial e}{\partial \rho} \right |_T +
         \left . \frac{\partial e}{\partial T} \right |_\rho  
         \left . \frac{\partial T}{\partial \rho} \right |_p 
         =
         \left . \frac{\partial e}{\partial \rho} \right |_T -
         \left . \frac{\partial e}{\partial T} \right |_\rho  
         \left . \frac{\partial p}{\partial \rho} \right |_T 
         \left ( \left . \frac{\partial p}{\partial T} \right |_\rho  \right )^{-1}
   \end{equation}
   where we used the constancy of pressure as:
   $dp = (\partial p/\partial \rho) |_T d\rho + (\partial p/\partial T) |_\rho dT = 0$
   to eliminate ${\partial T}/{\partial \rho} |_p$.

  Once we find $W_s$, we can evaluate $Z_s$ using CG Eqs.\ 20 and 23.  Two
  more thermodynamic derivatives are needed:
  \begin{align}
    \left . \frac{\partial p}{\partial e} \right |_\rho &=
    \left . \frac{\partial p}{\partial T} \right |_\rho 
    \left ( \left . \frac{\partial e}{\partial T} \right |_\rho \right )^{-1} \\
    \left . \frac{\partial p}{\partial \rho} \right |_e &=
    \left . \frac{\partial p}{\partial \rho} \right |_T -
    \left . \frac{\partial e}{\partial \rho} \right |_T 
    \left ( \left . \frac{\partial e}{\partial T} \right |_\rho \right )^{-1} \  \end{align}
  and in terms of specific volume, $\tau \equiv 1/\rho$, 
  \begin{equation}
  \left . \frac{\partial p}{\partial \tau} \right |_T = -\rho^2 \left .\frac{\partial p}{\partial \rho} \right |_T
  \end{equation}
  These can be derived in a similar fashion as shown above.

  We use the approximation for $W_s$ in the case where we know
  $\gamma_\star$ (CG, Eq. 24) as the initial guess for $W_s$.  We
  estimate a value for $\gamma_\star$ from the evolution equation CG
  derive, CG Eq.~31.

  \item {\em Rarefaction solution:}  In the case of a rarefaction, we 
  simply integrate the Riemann invariant ODEs.  We integate:
  \begin{equation}
  \frac{d\tau}{dp} = \begin{cases}
                        -{C^{-2}} & \text{left rarefaction} \\
                        -{C^{-2}} & \text{right rarefaction}
                     \end{cases}
  \qquad
  \frac{du}{dp}    = \begin{cases}
                        -{C^{-1}} & \text{left rarefaction} \\
                        \phantom{-}{C^{-1}} & \text{right ratefaction}
                     \end{cases}
  \end{equation}
  Here $C = C(\tau,p)$ is the Lagrangian sound speed.  The integration
  is done simply with 4-th order Runge Kutta from $p = p_s$ to $p_\star$.
  Then $W_s = C$ and $Z_s$ is given by CG Eq.~16.
 
 \end{itemize}

\item {\em Sampling the solution:}

 Once we have solved for the wave structure, we can sample the
 solution at a given time, $t$.  Given $N$ points, the coordinate
 centers are $x_i = (i + 1/2)\Delta x$ for $i = 1, \ldots, N$, with
 $\Delta x = (x_\mathrm{max} - x_\mathrm{min})/N$.  Then we define
 \begin{equation}
    \xi_i = \frac{x_i - \frac{1}{2}(x_\mathrm{max} - x_\mathrm{min})}{t}
 \end{equation}
 The speed of the contact $u_\star$ tells us which states we need to 
 deal with, and the evaluation of the solution proceeds from CG Eq.~15
 and following.  

 The only complication is the case where we are sampling
 inside the rarefaction.  As suggested in CG, we integrate the Riemann 
 invariants, however, we found that it is easiest to rewrite them
 with $u$ as the independent variable.  We integrate:
  \begin{equation}
  \frac{d\tau}{du} = \begin{cases}
         \phantom{-}{C^{-1}} & \text{left rarefaction} \\
                  -{C^{-1}}  & \text{right rarefaction} 
                     \end{cases}
  \qquad
  \frac{dp}{du}    = \begin{cases}
                    -{C}  &\text{left rarefaction} \\
                     \phantom{-}{C} & \text{right rarefaction}
                     \end{cases}
  \end{equation}
  We integrate from $u = u_s$ to
  \begin{equation}
  u_\mathrm{stop} = \begin{cases}
     \xi + c(\tau,p) & \text{left rarefaction} \\ 
     \xi - c(\tau,p) & \text{right rarefaction}
   \end{cases}
  \end{equation}
  We don't know $u_\mathrm{stop}$ at the start of the integration
  (since $c$ depends on the state), so we evaluate it each step.  Again
  we use Runge-Kutta integration, and we adjust our stepsize to stop at
  the converged $u_\mathrm{stop}$.

\end{itemize}

We use this solution to generate several exact shock tube profiles to
compare with the PPM solutions.  It is easier to specify the initial
conditions in terms of a jump in $(\rho, u, T)$ instead of $(\rho, u,
p)$ and then compute the pressure via the equation of state.
Additionally, for this equation of state, we need to specify the
composition of the gas---we take it to be pure $^{12}\mathrm{C}$.

Since even different versions of the same equation of state can have
slightly different thermodynamics, we do not provide a table of the
solution output for the problems listed here.  Instead, the complete
source code for this exact solver is available in the CASTRO source
distribution (in the \texttt{Util/exact\_riemann/} subdirectory).  This can be built with either the general stellar
equation of state \citep{timmesswesty} or a gamma-law
EOS.  All the necessary inputs files for the tests described here
are provided.

We devise 4 tests.  The first 3 are analogous to tests 1 through 3 in
\citet{toro:1997}: test~1 is a Sod-like problem, test~2 is a double rarefaction,
and test~3 is a strong shock.  The last test, test~4, mimics the conditions present at the
edge of a star when it is not very well resolved on a grid---a common
issue with multi-dimensional stellar simulations.
Table~\ref{table:tests} lists all the parameters, including the length
of the domain, $L$, and the end time of the simulation, $t$.  All tests
used outflow (zero-gradient) boundary conditions.  For test~4, because of the
large drop in density, the Coulomb corrections in the equation of state
were disabled to avoid unphysical regions in the thermodynamic space.

\clearpage


\clearpage

\begin{deluxetable}{lllcllcllcll}
\renewcommand{\arraystretch}{1.2}
\tablecolumns{9}
\tablewidth{0pt}
\tablecaption{\label{table:tests} Test problems.}
\tablehead{\colhead{quantity} &
           \multicolumn{2}{c}{test 1} & ~ &
           \multicolumn{2}{c}{test 2} & ~ &
           \multicolumn{2}{c}{test 3} & ~ &
           \multicolumn{2}{c}{test 4}
 \\
           \colhead{} & 
           \colhead{left} & \colhead{right} & ~ &
           \colhead{left} & \colhead{right} & ~ &
           \colhead{left} & \colhead{right} & ~ &
           \colhead{left} & \colhead{right}}
\startdata
$\rho~(\gcc)$     &
       $10^7$ & $10^6$ & ~ &  
       $10^7$ & $10^7$ & ~ &  
       $10^6$ & $10^6$ & ~ &  
       $10^2$ & $10^{-4}$ \\
$T~(\mathrm{K})$     &
       $10^8$ & $10^6$ & ~ &  
       $10^8$ & $10^8$ & ~ &  
       $10^9$ & $10^6$ & ~ &  
       $10^7$ & $10^7$ \\
$u~(\mathrm{cm/s})$     &
       $0$ & $0$ & ~ &  
       $-10^8$ & $10^8$ & ~ &  
       $0$ & $0$ & ~ &  
       $0$ & $0$ \\
$L~(\mathrm{cm})$    &
       \multicolumn{2}{c}{$10^6$} & ~ &   
       \multicolumn{2}{c}{$10^5$} & ~ &   
       \multicolumn{2}{c}{$2\times 10^5$} & ~ &   
       \multicolumn{2}{c}{$10^5$} \\   
$t~(\mathrm{s})$    &
       \multicolumn{2}{c}{$8\times 10^{-4}$} & ~ &   
       \multicolumn{2}{c}{$8\times 10^{-5}$} & ~ &   
       \multicolumn{2}{c}{$2\times 10^{-4}$} & ~ &   
       \multicolumn{2}{c}{$3\times 10^{-4}$} \\   
\enddata
\end{deluxetable}

\clearpage

\begin{figure}
\includegraphics[width=\linewidth]{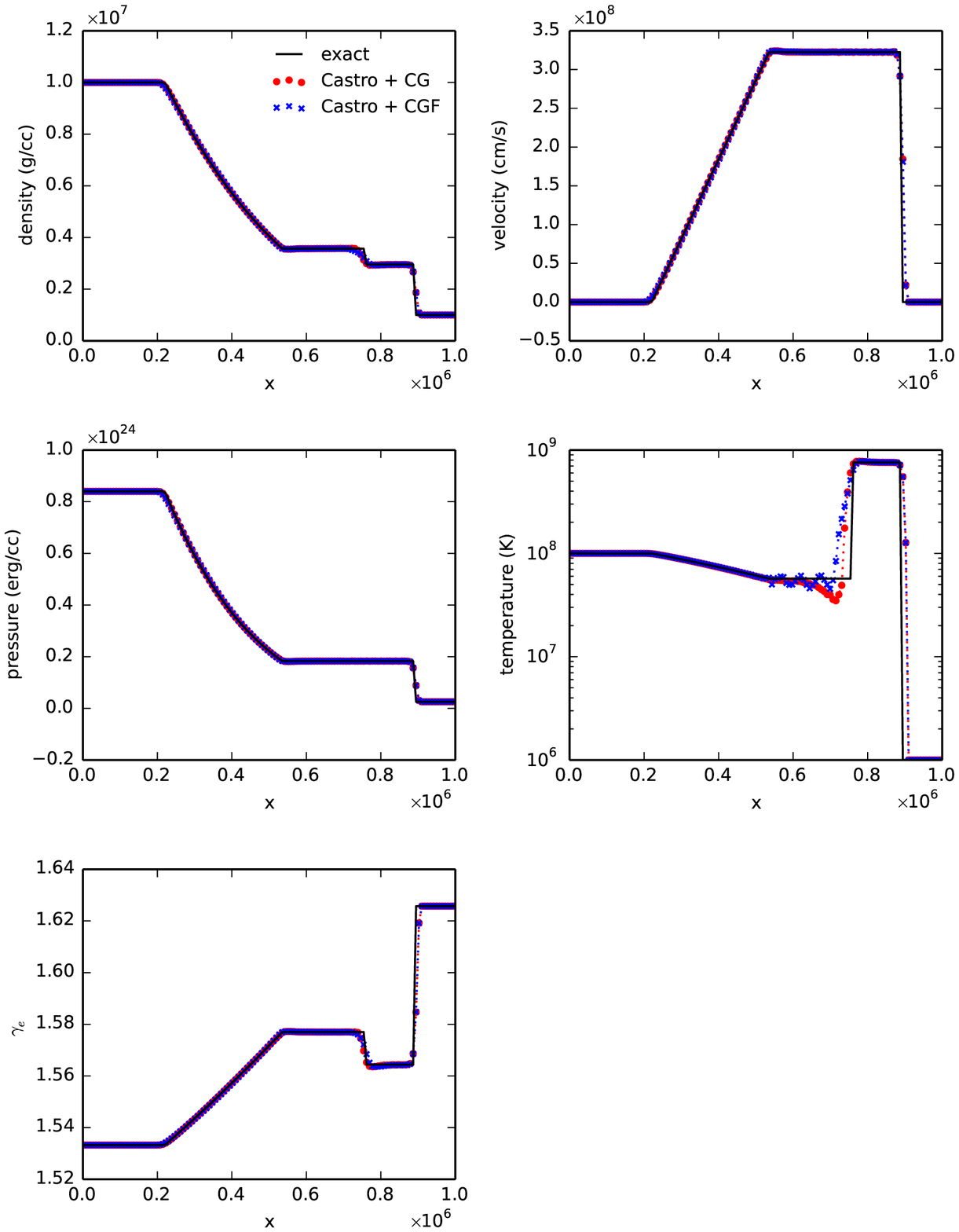}
\caption{\label{fig:riemann} Sod-like test 1 problem with the general EOS comparing the
differences between the CG and CGF Riemann solvers}
\end{figure}

\clearpage

\begin{figure}
\includegraphics[width=\linewidth]{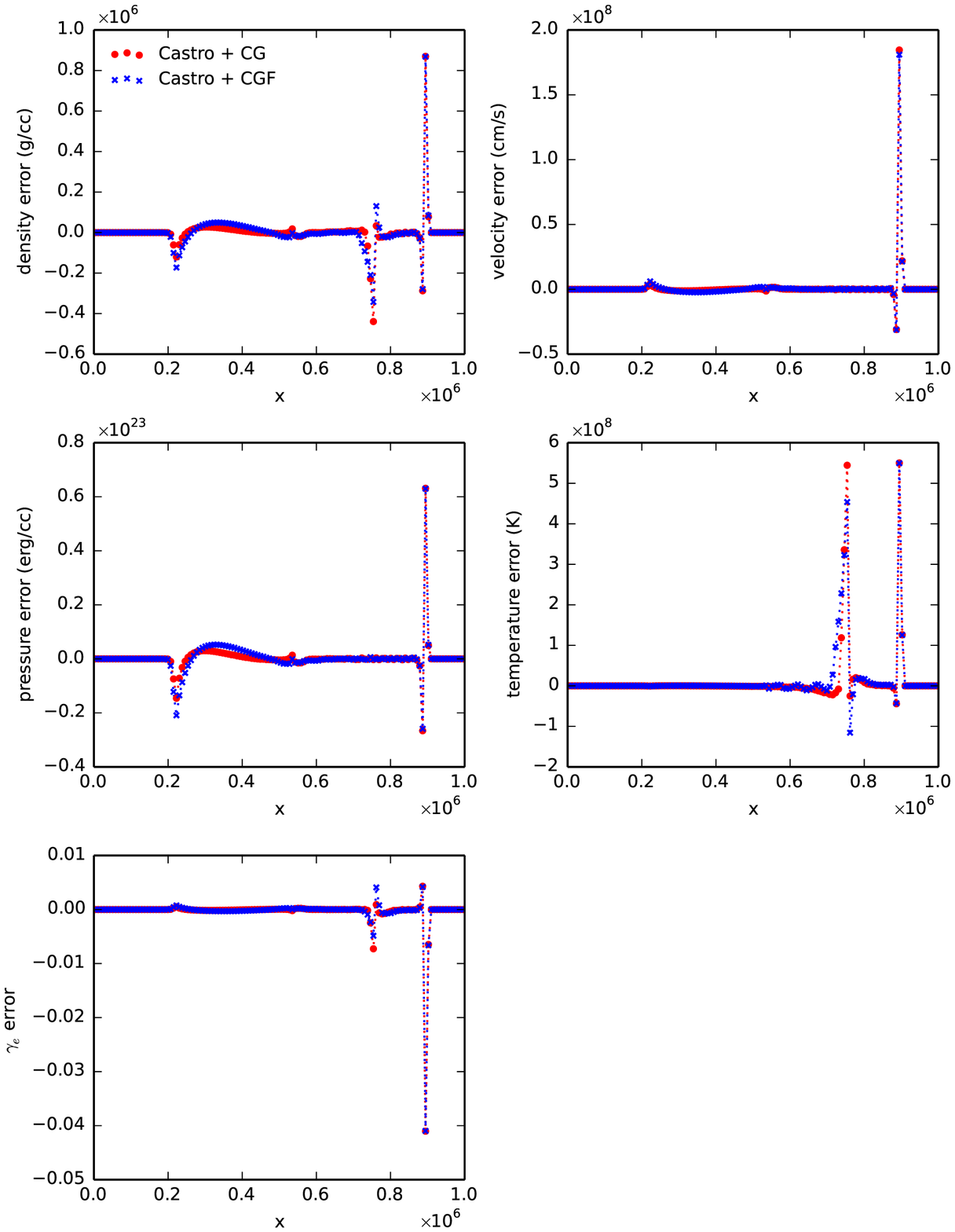}
\caption{\label{fig:riemann-error} Errors in the Sod-like test 1
  problem with the general EOS comparing the differences between the CG and CGF
  Riemann solvers}
\end{figure}

\clearpage

\begin{figure}
\includegraphics[width=\linewidth]{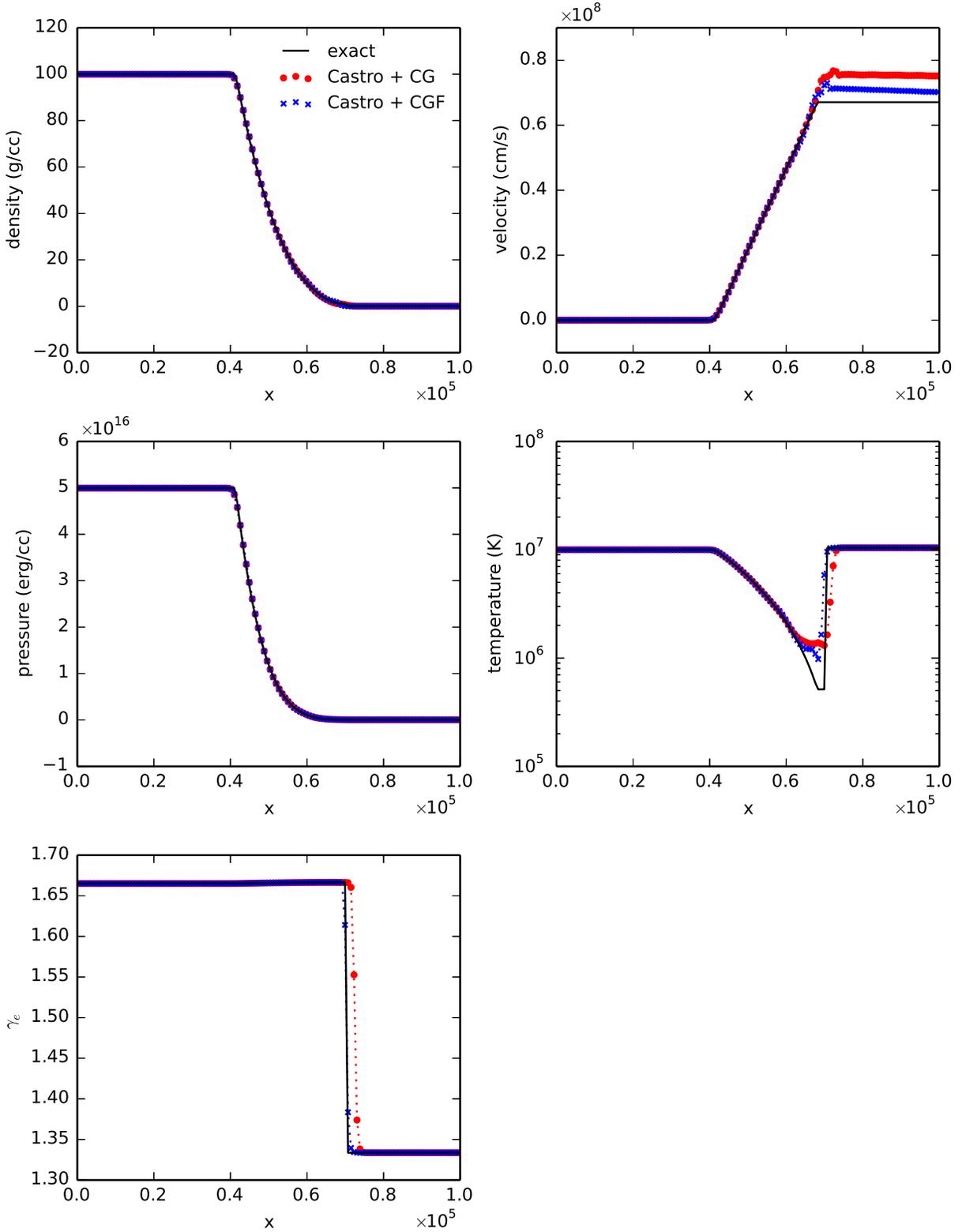}
\caption{\label{fig:riemann-error-test4} Solutions for the ``stellar
  edge'' test 4 problem with the general EOS comparing the differences between
  the CG and CGF Riemann solvers}
\end{figure}

\clearpage

\begin{figure}
\centering
\includegraphics[width=0.6\linewidth]{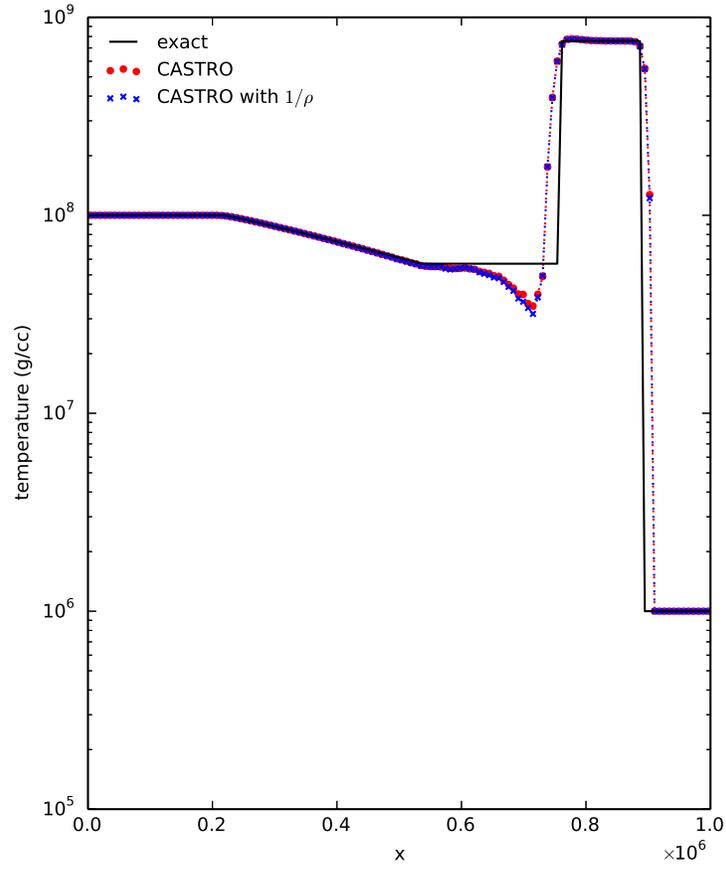}
\caption{\label{fig:test1-MC-CW} Sod-like test 1 problem with the
  general EOS comparing the use of $\tau = 1/\rho$ (CW) to $\rho$ (MC)
  in the characteristic tracing.  Only the temperature is shown}
\end{figure}

\clearpage

\begin{figure}
\includegraphics[width=\linewidth]{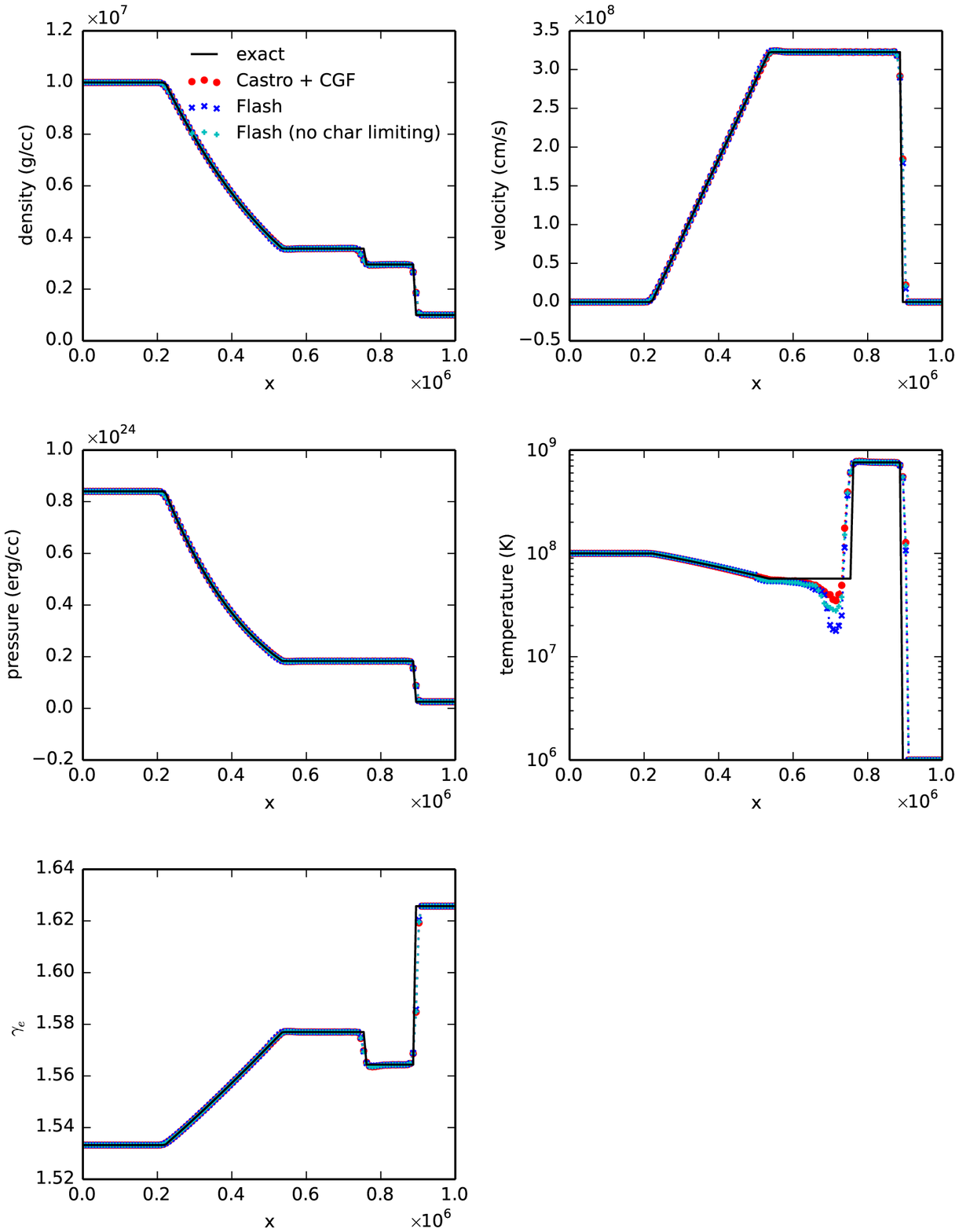}
\caption{\label{fig:summary-test1} A comparison of CASTRO to FLASH for
test 1.}
\end{figure}

\clearpage

\begin{figure}
\includegraphics[width=\linewidth]{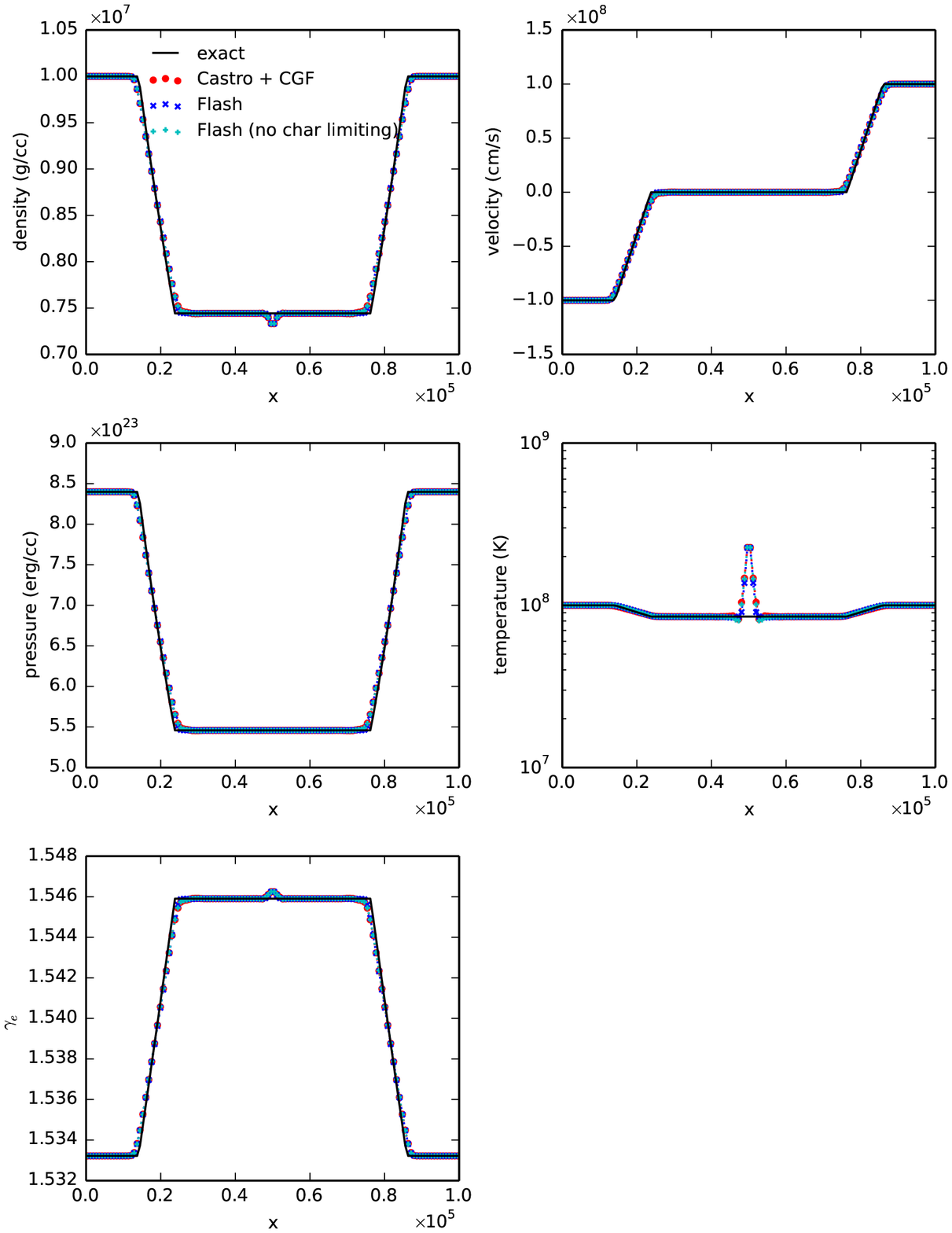}
\caption{\label{fig:summary-test2} A comparison of CASTRO to FLASH
for test 2.}
\end{figure}

\clearpage

\begin{figure}
\includegraphics[width=\linewidth]{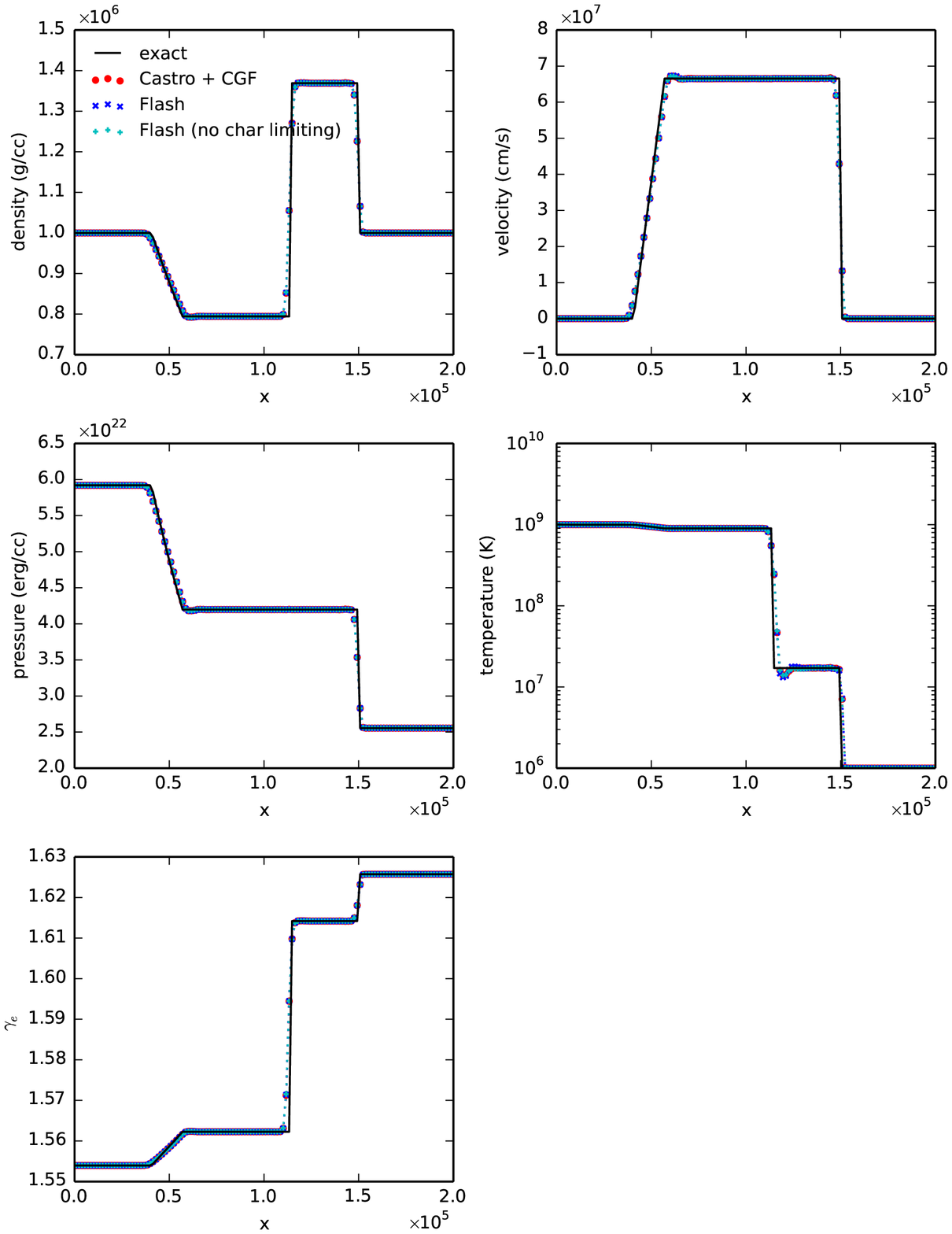}
\caption{\label{fig:summary-test3} A comparison of CASTRO to FLASH
to test 3.}
\end{figure}

\clearpage

\begin{figure}
\includegraphics[width=\linewidth]{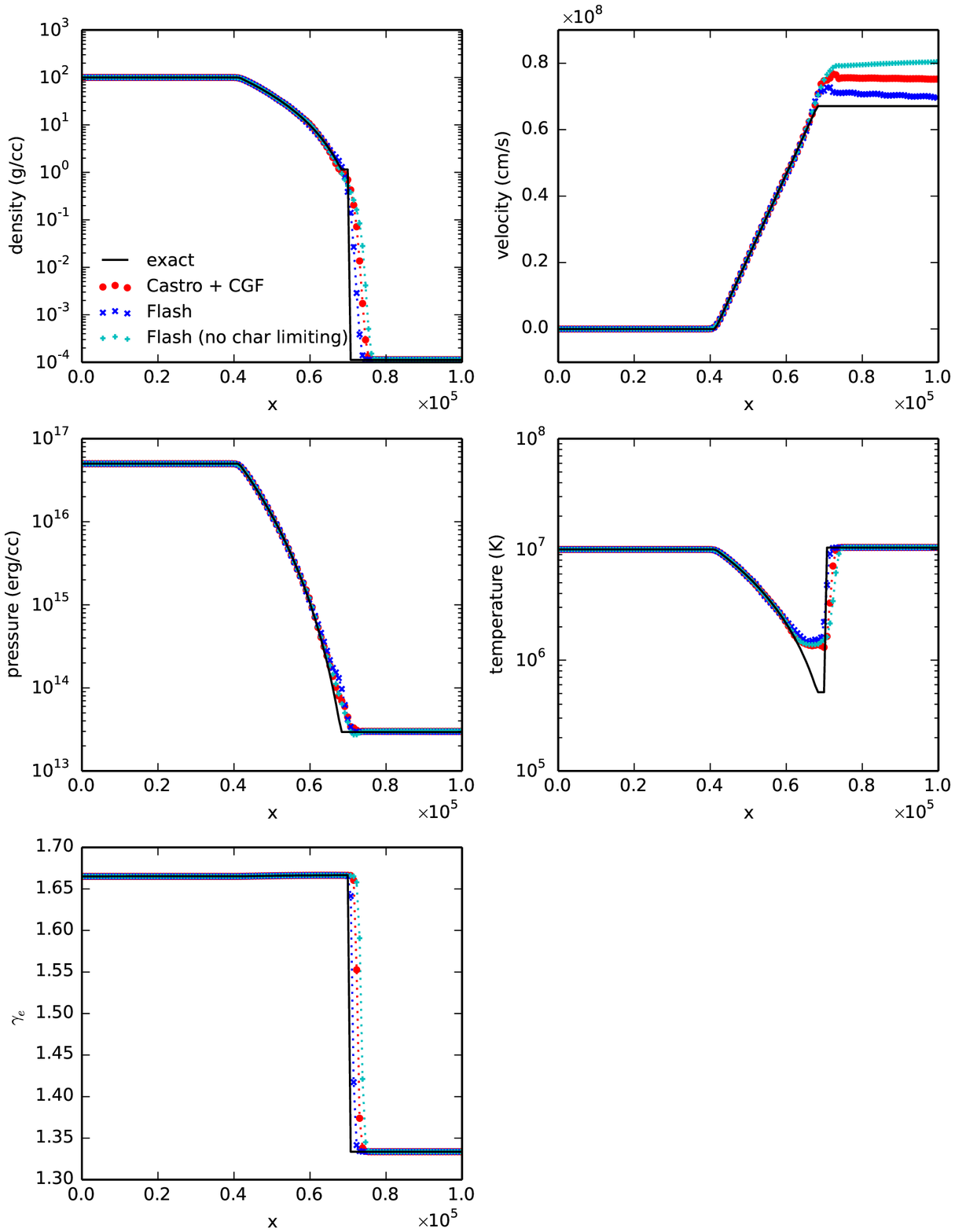}
\caption{\label{fig:summary-test4} A comparison of CASTRO to FLASH
to test 4.}
\end{figure}

\end{document}